\documentclass[12pt]{article}

\usepackage{amsmath,amssymb,array,amsfonts}
\usepackage[dvipdfmx]{graphicx}
\usepackage{comment,color}
\usepackage{ascmac}
\usepackage{cite}

\newcommand{\mot}{\!\not \!}
\newcommand{\mmot}{\!\!\not \!\! }

\newcommand{\ba}{\begin{eqnarray}}
\newcommand{\ea}{\end{eqnarray}}

\allowdisplaybreaks

\def\ncm{\newcommand}
\def\M {{\rm M}}
\def\e {{\rm e}}

\def\H {{\rm H}}
\def\B {{\rm B}}
\def\SM{{\rm SM}}

\def\rmd{{\rm d}}

\def\nt{\notag}

\ncm{\sls}[1]{{\ooalign{\hfil/\hfil\crcr$#1$}} }

\makeatletter
    
    \@addtoreset{equation}{section}
\makeatother

\begin{document}
\setlength{\baselineskip}{18pt}

\begin{titlepage}

\begin{flushright}
OCU-PHYS-486
\end{flushright}
\vspace{1.0cm}
\begin{center}
{\LARGE\bf 
Triple Higgs Boson Coupling \\
\vspace*{3mm}
in Gauge-Higgs Unification
} 
\end{center}
\vspace{25mm}

\begin{center}
{\Large
Yuki Adachi 
and 
Nobuhito Maru$^{*}$
}
\end{center}
\vspace{1cm}
\centerline{{\it
Department of Sciences, Matsue College of Technology,
Matsue 690-8518, Japan.}}

\centerline{{\it
$^{*}$Department of Mathematics and Physics, Osaka City University, Osaka 558-8585, Japan.
}}
%
%
\vspace{2cm}
\centerline{\large\bf Abstract}
\vspace{0.5cm}

We consider the triple coupling of the Higgs boson 
 in the context of the gauge-Higgs unification scenario.
We show that the triple coupling of the Higgs boson in this scenario 
 generically deviates from SM prediction
since the Higgs potential in this scenario has a periodicity.
We calculate the coupling in the five-dimensional $SU(3) \times U(1)_X$ gauge-Higgs unification model
 and obtain 70\% deviation from the SM prediction. 


\end{titlepage}

\newpage
\section{Introduction}

The standard model (SM) is almost established 
 by the discovery of the Higgs particle at the Large Hadron Collider (LHC) experiment.
The SM predictions 
are precisely measured by the several experiments 
 and highly consistent with each other.
Although the SM succeed in explaining various experimental data,
 the Higgs boson self-couplings are still unclear.
The future experiments (such as High Luminosity LHC (HL-LHC) or International Linear Collider (ILC))
 are expected to determine these couplings 
 and it therefore motivates us to investigate them.

The triple Higgs boson coupling $\lambda_{hhh}$ is defined 
 by the third derivative of the Higgs potential at the vacuum
\begin{equation}
\lambda_{hhh} =\left.  \frac{\partial^3 V}{\partial H^3}\right |_{H=v}
\end{equation}
where $v$ stands for the vacuum expectation value (VEV) of the Higgs boson field.
Noting that the Higgs potential of the SM is parameterized 
 by the mass parameter 
 and the quartic coupling, 
 we find
\begin{equation}
\lambda_{hhh}^{\SM}
=
\frac{3m_h^2}{v}
\end{equation}
where $m_h$ stands for the Higgs boson mass.

In the gauge-Higgs unification (GHU) scenario \cite{GH,HIL} where the SM Higgs boson is identified 
 with the extra spatial component of the gauge field in higher dimensions,
 the Higgs potential is induced by the quantum effects and expressed schematically as 
\begin{equation}
V(H)
=
 -\sum_{n=1}^\infty\sum_{k} a_k(n) \cos (\pi Rnk H)
\end{equation}
where $R$ stands for the radius of the $S^1$.
The index $k$ depends on the representations which the fields belong to.
Reflecting the higher dimensional gauge symmetry,
the Higgs potential has a periodicity as $V(H+2/R)=V(H)$. 
The coefficients $a_k(n)$ are basically proportional to $\frac{1}{n^5}$ in five-dimensional space-time. 
The Higgs boson mass and potential are finite.\footnote{For explicit loop calculations in various models, 
 see \cite{HIL, Finitemass}}
The Higgs boson parameters in this scenario are generically determined by 
\begin{align}
m_h^2=&\left.  \frac{\partial^2 V}{\partial H^2}\right |_{H=v},
\\
\lambda_{hhh}^\text{GHU}
=&\left. \frac{\partial^3 V}{\partial H^3}\right |_{H=v}. 
\end{align}
Combining these relations, we have
\begin{equation}
\lambda_{hhh}^{\rm GHU} = \frac{\partial m_h^2}{\partial v}.
\end{equation}
The above relation is different from the SM prediction.
It indicates that the triple coupling of the Higgs boson in the GHU scenario is expected to deviate from that of the SM,
 although the Higgs boson mass and the VEV are completely same as those in the SM.
It is very similar to the fermion masses in the GHU scenario\cite{Kurahashi:2014jca,Adachi:2015ova}, 
 which is also a periodic function of Higgs field VEV 
 contrary to the SM Yukawa couplings proportional to the VEV of Higgs field. 
Because of the periodic nature for Yukawa couplings, 
 the non-linear dependence of Higgs VEV on fermion masses is inevitable 
 and the deviation is anticipated.  
The similar reasoning can be hold true for the triple Higgs boson coupling in GHU.  
Following this observation, 
 we calculate the triple coupling of the Higgs boson in a model of GHU,
 which is proposed by the present authors in \cite{Adachi:2018mby}.

This paper is organized as follows. 
In section 2, we briefly describe our model. 
The Higgs potential calculation and the triple coupling of the Higgs boson are shown in the section 3. 
Summary is devoted to section 4.


\section{Model}
We consider the $SU(3) \times U(1)_X$ gauge theory in five-dimensional flat space-time.
The fifth extra dimension is compactified on an orbifold $S^1/Z_2$ where the radius of $S^1$ is $R$.
Since we have already analyze our model in detail \cite{Adachi:2018mby}, 
 we just describe here the outline of our model.\footnote{
The gauge sectors of our model has been discussed in detail \cite{Adachi:2016zdi}.
}

\begin{figure}
\begin{center}
\includegraphics{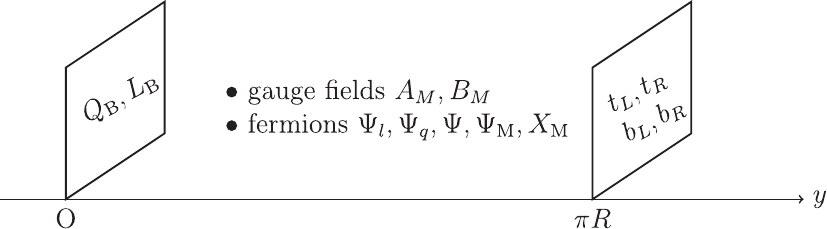}
\caption{Setup of the model.}
\label{fig:outline}
\end{center}
\end{figure}

The third generation of quarks are the brane-localized fermions
 which are put on the $y=\pi R$ brane. 
The other SM chiral fermions are the bulk fermions.
They are embedded in $\bf 3$ and $\overline{\bf 3}$ representations of the $SU(3)$.
We introduce further bulk fermions called as \lq\lq messenger fermion\rq\rq and \lq\lq mirror fermion\rq\rq .
The messenger fermions connect the third generation quarks and the Higgs boson fields ($A_y$)
 because the brane-localized fermions cannot interact with Higgs boson fields directly.
In order to reproduce the observed top quark mass, 
 we choose the $\bf\overline{15}$ representations of $SU(3)$ as the messenger fermion
 since the large representation can enhance the yukawa coupling 
 which is given by the gauge coupling in GHU. 
In order to realize the electroweak symmetry breaking (EWSB), 
 we need to introduce the mirror fermions propagating in the bulk.
They may be the possible candidate of the dark matter
 as pointed out in \cite{Maru:2017otg}.
In our model, 
 we choose the $\bf\overline{15}$ representation as the messenger fermions
 to obtain an appropriate EWSB.
The outline of this model is depicted in Fig. \ref{fig:outline}.

To summarize,
the Lagrangian for the fermion sectors are shown below.
For the third generation of the quarks,  it is given by  
\begin{align}
\mathcal L_\text{fermion}
\supset  &
\bar \psi ({\bf 3_0}) i\mmot D_{{\bf 3}}\psi ({\bf 3_0})
 +\bar \psi (\overline{\bf 15}_{\bf -2/3}) i\mmot D_{\overline{\bf 15}}\psi (\overline{\bf 15}_{\bf -2/3})
 \nonumber \\&
+\delta(y-\pi R)
\Big[
\bar Q_{L} i\partial_\mu\gamma^\mu Q_{L} 
+\bar t_R i\partial_\mu\gamma^\mu t_R
+\bar b_R i\partial_\mu\gamma^\mu b_R
\nonumber \\
&
+\frac{\epsilon_{L}}{\sqrt{\pi R}}Q_{L}
(\cos\theta Q_{15R}+\sin\theta Q_{3R})
+\frac{\epsilon_{tR}}{\sqrt{\pi R}} \bar T_L t_R
+\frac{\epsilon_{bR}}{\sqrt{\pi R}} \bar B_L b_R
+{\rm h.c.}
\Big]
\end{align}
where the $t_R, b_R$ and $Q_L$ are the chiral fermions of the third generation of the SM quarks.
The covariant derivative is $D^M=\partial^M+ig_5 A^M_a T^a+ig'_5 Q_X B^M~(M=0,1,2,3,4)$.
The $A^M$ and $B^M$ represent the gauge fields of $SU(3)$ and $U(1)_X$, respectively.
The numbers in the subscript of the fields and the covariant derivatives represent the each $U(1)_X$ charges.

As for other SM fermions, it becomes
\begin{align}
\mathcal L_\text{fermion}\supset&
\sum_{i=1}^2
\left\{
\bar\Psi^i_q({\bf 3_0}) \left[i\mot D_{\bf 3}+M_q^i\epsilon(y)\right] \Psi_q^i({\bf 3_0})
+\bar\Psi^i_q(\bar{\bf 3}_{\bf 1/3}) \left[i\mot D_{\bar{\bf 3}}+M_q^i\epsilon(y)\right] \Psi_q^i(\bar{\bf 3}_{\bf 1/3})
\right\} \nonumber 
\\
&\nt
+
\sum_{i=1}^3
\left\{\bar\Psi^i_l({\bf 3}_{\bf -2/3}) \left[i\mot D_{\bf 3}+M^i_l\epsilon(y)\right] \Psi^i_l({\bf 3}_{\bf -2/3})
+\bar\Psi^i_l(\bar{\bf 3}_{\bf -1/3}) \left[i\mot D_{\bar{\bf 3}}+M^i_l\epsilon(y)\right] \Psi^i_l(\bar{\bf 3}_{\bf -1/3})
\right\}
\nonumber 
\\ \nt 
&+ \delta(y)
\sum_{i=1}^2
 \left[
  \bar Q_\B^i i\partial_\mu\gamma^\mu Q^i_\B
  +\frac{\epsilon_q}{\sqrt{\pi R}}
  \bar Q^i_{\B}Q^i_{\rm H} +{\rm h.c.}
 \right]
 \\
&+ \delta(y)
\sum_{i=1}^3
 \left[
  \bar L_\B^i i\partial_\mu\gamma^\mu L^i_\B
  +\frac{\epsilon_l}{\sqrt{\pi R}}
  \bar L^i_{\B}L^i_{\rm H} +{\rm h.c.}
 \right]. 
\end{align}
The $\Psi_q({\bf 3_0})$ includes the massless chiral fermions 
 of the down-type right-handed quark singlets and the left-handed quark doublets.
Similarly, the $\Psi_q(\bar{\bf 3}_{\bf 1/3})$ includes 
 the massless chiral fermions of the up-type right-handed quark singlets 
 and the left-handed quark doublets.
The $Q_\B$ and $L_\B$ are the brane-localized fermions that couples to the bulk fermions.
The $Q_\H$ and $L_\H$ are a mixture of the doublets of the $\bf 3$ and $\bf\overline 3$ representations.
In this setup, 
 we note that two sets of quark and lepton doublets are present per generation.
Therefore, the redundant doublets are made massive 
 and removed from the low energy effective theory by these brane-localized interactions.\footnote
 {These brane-localized interactions are also important to reproduce the flavor mixing \cite{flavorGHU}.}

The Lagrangian of the mirror fermions $\Psi_\M$ and $X_\M$ is
\begin{equation}
\mathcal L_\text{fermion}
\supset 
\bar\Psi_\M i\mot D_{\bf 15} \Psi_\M
+\bar X_\M i\mot D_{\bf 15} X_\M
+M
\left[\bar\Psi_\M X_\M +\bar X_\M\Psi_\M \right].
\end{equation}
Note that 
they can have a constant mass 
 since the opposite $Z_2$ parities will be assigned to the mirror fermion as in the next paragraph.
The massless modes in the mirror fermions become massive by this bulk mass term.

We impose the periodic boundary conditions on the bulk fermions except for the messenger fermions 
 and their $Z_2$ parity as follows;
\begin{equation}
\Psi(y+2\pi R) = \Psi(y), \quad 
\Psi(y)=-P_{{\bf \overline{15}} } \gamma_5 \Psi (-y) 
\end{equation}
where the $P_{{\bf \overline{15}} }$ is defined as the tensor product of  
$P_{{\bf 3}}=\text{diag}(+,+,-)$ for the $SU(3)$ fundamental representation.
Then, the remaining chiral fermions in the $\bf 3$ and $\bf \bar{3}$ representations become the SM chiral fermions.
The boundary conditions for the mirror fermions are given by
\begin{align}
\label{BC_mirror1}
\Psi_\M(+ y)=P_{{\bf 15}} \gamma_5 \Psi_\M(- y), \quad
\Psi_\M(y)=\Psi_\M(y+2\pi R),
\\
X_\M(+ y)=-P_{{\bf 15}} \gamma_5 X_\M(- y), \quad 
X_\M(y)=X_\M(y+2\pi R).
\end{align}
Note that the assigned $Z_2$ parities are opposite each other as mentioned above.

As for the messenger fermions, 
 the anti-periodic boundary conditions and the following $Z_2$ parity are imposed.
\begin{equation}
\psi(y+2\pi R) = -\psi(y), \quad 
\psi(y)=-P\gamma_5 \psi (-y),
\end{equation}
where $P$ represents $P_{{\bf 3}}(P_{\bf 15})$ for the ${\bf 3}({\bf 15})$ representation of $SU(3)$, respectively. 
Adopting such boundary conditions simplifies our model,
 namely, the massless chiral fermions in the messenger fermions 
 are completely projected out without introducing the complicated boundary terms.


\section{Effective potential and the triple coupling of the Higgs boson field} 

From the Lagrangian that we have shown in the above section,
 the mass spectrum of this model is obtained.
Since we have already derived them in detail \cite{Adachi:2018mby},
we only show the results.
The top and bottom quarks have a mass $m_n$ satisfying the following equations.
\begin{align}
\label{massTOP}
0=&
 2\hat m_n^2\cos^2\hat m_n\left(\cos^2\hat m_n-\sin^2(2\hat M_W)\right)
 \left(\cos^2\hat m_n-\sin^2(4\hat M_W)\right)
\nonumber \\
&-\epsilon_L^2\hat m_n\cos \hat m_n\sin \hat m_n
\nonumber \\
&
 \times  \Big[
 \sin^2\theta\left\{\sin^2(4 \hat M_W) \cos^2\hat m_n 
 +\sin^2 (2 \hat M_W) \cos^2 \hat m_n
 -2\sin^2 (2\hat M_W) \sin^2 (4 \hat M_W)
 \right\}
 \nonumber \\
 &
 -2\cos^4\hat m_n+\sin^2(4\hat M_W) \cos^2\hat m_n
 +\sin^2(2\hat M_W) \cos^2 \hat m_n
 \Big]
\nonumber \\
&
 +\frac{\epsilon_{tR}^2}{4}\hat m_n\sin \hat m_n\cos \hat m_n
\nonumber  \\
&
 \times \Big[
 8\cos^4\hat m_n
 -7\sin^2(4\hat M_W)\cos^2\hat m_n
 -4\sin^2(2\hat M_W)\cos^2\hat m_n
 +3\sin^2(2\hat M_W)\sin^2(4\hat M_W)
 \Big]
\nonumber  \\
&-\frac{\epsilon_L^2\epsilon_{tR}^2}{8}
\Big[
 \cos^2\theta
 \Big\{
  8(\sin^2(4\hat M_W)+\sin^2(2\hat M_W))\cos^4\hat m_n
 \nonumber \\
  &
  +\cos^2\hat m_n
  \left(
    (-11\sin^2(2\hat M_W)-7)\sin^2(4\hat M_W)
    +\sin(4\hat M_W)\sin(8\hat M_W)
    -4\sin^2(2\hat M_W)\right)
  \nonumber \\
  &+6\sin^2(2\hat M_W)\sin^2(4\hat M_W)
 \Big\}
 +16\cos^6\hat m_n
 -2(7\sin^2(4\hat M_W)+4\sin^2(2\hat M_W)+8)\cos^4\hat m_n
 \nonumber \\
 &~~
 +2\cos^2\hat m_n
 \left\{
 (3\sin^2(2\hat M_W)+7)\sin^2(4\hat M_W)+4\sin^2(2\hat M_W)
 \right\}
 -6\sin^2(2\hat M_W)\sin^2(4\hat M_W)
\Big] 
\end{align}
and 
\begin{align}
\label{massBOTTOM}
0=&
-2\hat m_n^2(\sin^2\hat m_n-\cos^2\hat M_W)^2(\sin^2\hat m_n-\cos^2(3\hat M_W)) 
\nonumber \\
&
-\frac{\epsilon_L^2}{2} \hat m_n\sin\hat m_n\cos\hat m_n
(\sin^2\hat m_n-\cos^2\hat M_W)
\nonumber \\
&
\times \left[
\sin^2\theta(\cos^2(3\hat M_W)-\cos^2\hat M_W)
-4\sin^2(\hat m_n)+3\cos^2(3\hat M_W)+\cos^2(\hat M_W)
\right]
\nonumber \\
&
+2\epsilon_{bR}^2\hat m_n\sin\hat m_n\cos\hat m_n
 (\sin^2\hat m_n-\cos^2\hat M_W)
 (\sin^2\hat m_n-\cos^2(3\hat M_W))
 \nonumber \\
&
+\frac{\epsilon_L^2\epsilon_{bR}^2}{2}
\Big[
\Big\{(\cos^2\hat M_W-\cos^2(3\hat M_W))\sin^4\hat m_n 
-4\sin^2\hat M_W\cos^2\hat M_W\cos^2(3\hat M_W)
\nonumber \\
&
+ (\cos^2(3\hat M_W)-4\cos^4\hat M_W+3\cos^2\hat M_W)\sin^2\hat m_n
\Big\}\sin^2\theta
+4\sin^6\hat m_n 
\nonumber \\
&
- \left(3\cos^2(3\hat M_W)+\cos^2\hat M_W+4\right)\sin^4\hat m_n+(3\cos^2(3\hat M_W)+\cos^2(\hat M_W))\sin^2\hat m_n
\Big]
\end{align}
where $\hat m_n=\pi Rm_n$ and $\hat M_W=\pi R M_W$ are 
 dimensionless Kaluza-Klein (KK) mass and the W-boson mass normalized by $\pi R$, respectively.

We note that the exotic fermions with the different quantum numbers 
 from those of SM particle are included in the $\overline{\bf 15}$ representation.
Their spectrum are given by the solutions of the following equations
\begin{align}
\label{massexotic}
&0=\cos \hat m_n \cos(\hat m_n-2 \hat M_W) \cos( \hat m_n + 2 \hat M_W),
\nonumber \\
&0=\cos(\hat m_n- \hat M_W) \cos(\hat m_n+ \hat M_W),
\\
&0=\cos \hat m_n. 
\nonumber 
\end{align}
The lightest mode of these exotic fermions obtain a mass around $\sim 1/(2R)$
 due to the anti-periodic boundary conditions.

The down-type quarks have a mass $m_n$ that satisfy the equation;
\begin{align}
0&=
\sqrt{\hat m_n^2-\hat M_q^2}\epsilon^2_q
\left[(\hat M_q^2-\hat m_n^2)\sin^2\hat M_W \cos^2\sqrt{\hat m_n^2-\hat M_q^2}\sin^2\theta_q
+\hat m_n^2
\sin^2\sqrt{\hat m_n^2-\hat M_q^2}\right]
\nonumber \\
&\times \cos\sqrt{\hat m_n^2-\hat M_q^2}
 +\epsilon^2_q\sin^2\theta_q\hat M_q(\hat m_n^2-\hat M_q^2)\sin^2\hat M_W\sin\sqrt{\hat m_n^2-\hat M_q^2}
\nonumber \\&
 +\hat m_n^2\left[(\hat M_q\epsilon^2_q-2\hat m_n^2)\cos^2\sqrt{\hat m_n^2-\hat M_q^2}
 +2\hat m_n^2\cos^2\hat M_W+\hat M_q(2\hat M_q\sin^2\hat M_W-\epsilon^2_q)\right]
\nonumber \\
&\times \sin\sqrt{\hat m_n^2-\hat M_q^2}.
\end{align}
where the $M_q$ stands for the bulk mass for the quark.
As for the up-type quark,
 the KK mass spectrum can be found by solving a equation: 
\begin{align}
0=&
\sqrt{\hat m_n^2-\hat M_q^2}\epsilon^2_q\left[(\hat M_q^2-\hat m_n^2)\sin^2\hat M_W\cos^2\theta_q+\hat m_n^2\sin^2\sqrt{\hat m_n^2-\hat M_q^2}\right]\cos\sqrt{\hat m_n^2-\hat M_q^2}
\nonumber \\&
 +\hat M_q\epsilon_q^2(\hat m_n^2-\hat M_q^2)\sin^2\hat M_W\sin\sqrt{\hat m_n^2-\hat M_q^2}\cos^2\theta_q
\nonumber \\&
 +\hat m_n^2(\hat m_n^2-\hat M_q\epsilon_q^2)\sin^3\sqrt{\hat m_n^2-\hat M_q^2}
 +2\hat m_n^2(\hat M_q^2-\hat m_n^2)\sin^2\hat M_W\sin\sqrt{\hat m_n^2-\hat M_q^2}.
\end{align}

The KK mass spectrum of the mirror fermions is obtained 
 by solving the following equation;
\begin{equation}
0
=\sin\left(\sqrt{\hat m_n^2-\hat M^2}-\hat M_W\right)
\sin\left(\sqrt{\hat m_n^2-\hat M^2}+\hat M_W\right). 
\end{equation}
%
Noting that the bulk mass for the mirror fermions are constrained 
 from the search for the fourth generation fermions, 
 the mass of the lightest mode in the mirror fermion
 should be larger than the $\mathcal O(700 \rm GeV)$ or so \cite{Patrignani:2016xqp}, 
 which implies that the bulk mass of the mirror fermion
 must satisfy the lower bound
\begin{equation}
M_q>\sqrt{(700{\rm GeV})^2-M_W^2}. 
\end{equation}
The lepton sector is completely same as the quark,
their mass spectrum is obtained by replacing 
$M_q\to M_l,\theta_q\to \theta_l$ and $\epsilon_q\to \epsilon_l$.

Finally, we comment on the the $W$ and $Z$ bosons.
Their KK mass spectrum, which are precisely argued in \cite{Adachi:2016zdi},
are given by solving the following equations;
\begin{align}
0=& \cos^2 (\hat m_n) - \cos^2(\hat M_W)&\text{for the $W$ boson},
\\
0
=&
\tan^2(\hat m_n) -
\frac{\sin^2(\hat M_W)[4\cos^2\theta_W-\sin^2(\hat M_W)]}
{(2\cos^2\theta_W-\sin^2(\hat M_W))^2}&\text{for the $Z$ boson}.
\end{align}

Now, we calculate the effective potential for the Higgs boson field.
Generically, a particle with the mass $m_n$ contributes to the 1-loop effective potential as follows. 
\begin{equation}
 \label{EP1}
 V_\text{5D}
 =
\frac{1}{2\pi R} 
\frac{(-1)^F N_\text{DOF}}{2}
\int\frac{\rmd ^4p_{\rm E}}{(2\pi)^4}
 \sum_{n=-\infty}^{\infty}
 \ln\left(p_{\rm E}^2+m_n^2\right)
\end{equation}
where the $N_\text{DOF}$  stands for the degree of freedom of the fields running in the loop 
 and $F=1(0)$ is a fermion number for the fermion (boson).
The above expression (\ref{EP1}) can be rewritten by the following integral form as
\begin{align}
 \label{EP_integral_form}
  V_\text{5D} 
 =&
 -\frac{1}{2\pi R} 
 \frac{(-1)^FN_\text{DOF}}{32\pi^2}
 \frac{1}{R^4}
 \int_0^\infty \rmd u~u^{4}\frac{\rmd}{\rmd u}\ln\left[N(iu)\right]. 
\end{align}
The mass spectrum $m_n$ is determined by zeros of the function $N(iu)$, 
\begin{equation}
N(m_n)=0.
\end{equation}
The function $N(iu)$ is defined as such that $\hat m_n$ and $\hat M_W$ are replaced 
 by $i \pi u$ and $\pi \alpha$ in the equations determining the KK mass spectrum, respectively. 
Then, the four dimensional effective potential in our model is described as follows;
\begin{equation}
V_\text{4D~GHU} = \int_0^{2\pi R}\rmd y~ V_\text{5D}
=- 
 \frac{(-1)^FN_\text{DOF}}{32\pi^2}
 \frac{1}{R^4}
 \int_0^\infty \rmd u~u^{4}\frac{\rmd}{\rmd u}\ln\left[N(iu)\right]. 
\end{equation}
Finally, the 1-loop Higgs boson effective potential of our model is given by 
\begin{align}
V_\text{4D~GHU}
=&
 -
 \frac{1}{32\pi^2}
 \frac{1}{R^4}
 \int_0^\infty \rmd u\,u^{4}
 \frac{\rmd}{\rmd u}
 \Big[
 3\ln N_Z(iu)+3\ln N_W(iu)
 \nonumber \\
 &
 -3\cdot4\ln N_\text{BOT}(iu)
 -3\cdot4\ln N_\text{TOP}(iu)
 -3\cdot4\ln N_\text{exotic}(iu)
 -3\cdot4\ln N_\text{M}(iu)
 \Big] 
 \nonumber \\
 & -(\alpha\to 0)
  \label{effpot}
\end{align}
where $N_{Z}(iu)$, $N_{W}(iu)$, $N_\text{BOT}(iu)$, $N_\text{TOP}(iu)$, $N_\text{exotic}(iu)$ and $N_\text{M}(iu)$ 
 are the functions which determine the KK mass spectrum 
 for the $Z$ boson, $W$ boson, the bottom quark, top quark, the exotic fermions and the mirror fermions, respectively.


The triple coupling for the Higgs boson field $\lambda_{hhh}$ 
 is defined by the third derivative of the Higgs potential
 as shown in the introduction.
\begin{align}
\lambda_{hhh}^\text{GHU}
=&\left. \frac{\partial^3 V_\text{4D~GHU}}{\partial H^3}\right |_{H=v}
\end{align}
In the  SM, the triple coupling of the Higgs boson is given by
\begin{equation}
\lambda_{hhh}^\SM
 =3m_{h}^2 \frac{1}{v}
\sim 0.191~\rm TeV
,
\end{equation}
where we use $m_h=125~\rm GeV$ and $v = 246~\rm GeV$.
In our model, it is found 
\begin{equation}
\lambda_{hhh}^\text{GHU}
=\left. \frac{\partial^3 V_\text{4D~GHU}(H)}{\partial H^3}\right|_{H=v}
=\left. \frac{g^3}{8M_\text{KK}^3}\frac{\partial^3 V_\text{4D~GHU}(\alpha)}{\partial \alpha^3}\right|_{\text{VEV}}
=
0.334~\rm TeV,
\end{equation}
which is 70\% deviation from the SM prediction. 
We expect that the future experiments such as HL-LHC or ILC will detect the deviation.

We discuss the validity of the relatively large deviation obtained above.  
First, let us note that the Higgs triple coupling and the Higgs mass 
 in the SM satisfy the following relation:
\begin{equation}
 \lambda_{hhh}^{\text{SM}}(v) = 3 \frac{m_h^2(v)}{v}.
\end{equation}
On the other hands, the above relation is not valid in the GHU scenario
 since it breaks the periodicity of the Higgs potential.
Then it suggests that the Higgs triple couplings in the GHU scenario should deviate from the SM one 
 although the Higgs mass and the vev are completely same as those in the SM.

The Higgs potential in the GHU scenario has the periodicity,
 which has a form as 
\begin{equation}
 V(H)=-\sum_{n=1}^\infty \sum_{k} a_k(n) \cos (\pi R nk H), 
\end{equation}
where the index $k$ depends on the representations which the fields belong to. 
In the case of five dimensions, 
 the coefficients $a_{k}(n)$ proportional to $1/n^5$ for the gauge bosons, 
 the massless fermions and mirror fermions with constant mass.
As for the fermions with $Z_2$ odd bulk mass term, 
 the coefficients contains an additional factor $\e^{-\pi RMn}$.

The vev $v$ is defined by solving a stationary condition 
\begin{equation}
0=\left.\frac{\partial V}{\partial H}\right|_{H=v}
=
 \sum_{n,k} a_k(n) (\pi Rnk)  \sin(\pi Rnkv). 
\end{equation}
The Higgs mass and the triple Higgs coupling are given by the derivative of the Higgs potential
\begin{align}
m^2_h
=&
\left.\frac{\partial^2 V}{\partial H^2}\right|_{H=v}
=
 \sum_{n,k} a_k(n) (\pi R nk)^2 \cos(\pi Rnkv),
\\
\lambda_{hhh}^\text{GHU}
=&
\left. \frac{\partial^3 V}{\partial H^3}\right |_{H=v}
=
 -\sum_{n,k} a_k(n) (\pi R nk)^3 \sin(\pi Rnkv). 
\end{align}
The parameters are chosen to reproduce the Higgs mass $m_h$ and the vev $v$ of the SM.
The deviation of the Higgs triple couplings from the SM prediction is obtained by
\begin{equation}
 \Delta \lambda_3
 =\lambda_{hhh}^\text{GHU}-\lambda_{hhh}^\text{SM}
 =\lambda_{hhh}^\text{GHU}-\frac{3m _h^2}{v}.
\end{equation}
Combining above relations, 
we have
\begin{align}
 \Delta \lambda_3
 =&
 -\sum_{n,k}
 a_k(n)(\pi Rnk)^2
 \left[ (\pi Rnk) \sin(\pi Rnmv) + \frac{3}{v} \cos (\pi Rnkv) \right]
 \\
 =&
 -\sum_{n,k}
 a_k(n)(\pi Rnk)^2
 {\sqrt{(\pi Rnk)^2+\frac{9}{v^2}} }
 \cos(\pi Rnk v +\beta). 
\end{align}
The $\beta$ in the above expressions stands for the phases.
Since the coefficients $a_k(n)$ are proportional to $\frac{1}{n^5}\frac{1}{(10 R)^4}$ or $\frac{1}{(10 R)^4}\e^{-\pi RMn}$,
 the infinite summation can be approximated as
\begin{align}
 \Delta \lambda_3
 \sim&
 -\sum_{k}
 \frac{1}{(10 R)^4}(\pi Rk)^2
 {\sqrt{(\pi Rk)^2+\frac{9}{v^2}} }
 \cos(\pi Rk v +\beta)
 \\
 \sim&
 -\sum_{k}
 \frac{1}{(10 R)^4}(\pi Rk)^3
\sim
 -\sum_k
 (10^{-1}\pi k)^3 (10^{-1} M_{\rm KK})
\end{align}
where $M_{\rm KK}=1/R$ is the KK scale. 

Noting that the most dominant contribution to the 
triple Higgs couplings 
comes form the fermion in large representation, 
it corresponds to $k=4$ in our model. 
Taking into account that the KK scale is of order TeV scale \cite{Adachi:2018mby}, 
 we can estimate the deviation of the Higgs triple coupling $\Delta \lambda_3$ 
 to be ${\cal O}(1)$ times the weak scale 
 unless an accidental cancellation among various terms in the Higgs potential. 
This observation supports our result.  

Furthermore, a large deviation of the Higgs triple coupling from the SM prediction 
 has been reported in other models \cite{GSW}. 
In this paper, the Higgs potential up to dimension six terms was considered 
and the deviation of Higgs triple coupling was calculated 
in terms of the function of Higgs mass and the cutoff scale.   
According to their results, 
around 60\% deviation was found for the case of 125GeV Higgs mass and the cutoff scale of TeV scale.  
In our Higgs potential, not only dimension six terms but also higher order dimension terms are included. 
Therefore, it seems to be natural to obtain similar results since both Higgs potentials are extended by adding the non-renormalizable terms.


\section{Summary}

In this paper, 
 we analyzed the triple coupling of the Higgs boson 
 within the five-dimensional $SU(3) \times U(1)_X$ GHU 
 which we have proposed in \cite{Adachi:2018mby} 
 where a successful electroweak symmetry breaking occurs in a simple matter content.
The triple coupling of the Higgs boson in the GHU scenario
 is possible to deviate from SM prediction generically 
since the Higgs potential has the periodicity 
 reflecting the higher dimensional gauge symmetry.
We calculated this coupling in this model
 and obtained the 70\% deviation from the SM prediction.
We expect that the deviation will be detected in a future HL-LHC or ILC experiments 
 and the validity of GHU will be verified. 


\section*{Acknowledgments}
The work of N.M. is supported in part by JSPS KAKENHI Grant Number JP17K05420.




\end{document}